\documentclass[twocolumn,aps,prb]{revtex4}

\usepackage{epsfig,graphicx}
\usepackage{amsmath,verbatim}

\begin{document}

\title{Growth Structure and Work Function of Bilayer Graphene on Pd(111)}

\author{Y. Murata$^1$, S. Nie$^{2}$, A. Ebnonnasir$^{3}$, E. Starodub$^{2}$, B.~B. Kappes$^{3}$, K.~F. McCarty$^{2}$, C.~V. Ciobanu$^{3*}$,
and S. Kodambaka$^{1}$\footnote{To whom correspondence may be
addressed, emails: kodambaka@ucla.edu, cciobanu@mines.edu}}

\affiliation{$^1$Department of Materials Science and Engineering, University of California Los Angeles, Los Angeles, CA 90095\\
$^2$Sandia National Laboratories, Livermore, CA 94550\\
$^3$Department of Mechanical Engineering and Materials Science
Program, Colorado School of Mines, Golden, Colorado 80401}

\begin{abstract}
Using {\em in situ} low-energy electron microscopy and density
functional theory, we studied the growth structure and work function
of bilayer graphene on Pd(111). Low-energy electron diffraction
analysis established that the two graphene layers have multiple
rotational orientations relative to each other and the substrate
plane. We observed heterogeneous nucleation and simultaneous growth
of multiple, faceted layers prior to the completion of second layer.
We propose that the facetted shapes are due to the zigzag-terminated
edges bounding graphene layers growing under the larger overlying
layers. We also found that the work functions of bilayer graphene
domains are higher than those of monolayer graphene, and depend
sensitively on the orientations of both layers with respect to the
substrate. Based on first-principles simulations, we attribute this
behavior to oppositely oriented electrostatic dipoles at the
graphene/Pd and graphene/graphene interfaces, whose strengths depend
on the orientations of the two graphene layers.
\end{abstract}

\maketitle


\section{Introduction}
Few-layer graphene\cite{Geim2007} is attractive for applications in
optoelectronics as transparent conductors --where precise control
over layer thickness is not necessary-- owing to its low sheet
resistance and high transparency,\cite{Kim2009,{Bae2010}} and in
field-effect transistors due to the opening of an electronic bandgap
that, at least in bilayer graphene, can be tuned with an electric
field.\cite{Zhang2009} Given that device characteristics depend on
the work function of graphene and on how it varies at
contacts,\cite{Xia2011} it is of fundamental and practical
importance to understand the nature of the contact (Ohmic or
Schottky) at metal-graphene interfaces. In the case of monolayer
graphene, recent studies indicate that the crystallographic
orientation of graphene domains with respect to the metal can alter
their electronic
properties.\cite{Kwon2009,Murata2010APL,Murata2010ACSNano} For
few-layer graphene, the role of the specific metal surface and of
the in-plane orientations of those layers on the electrical
characteristics of the graphene is not well known. We choose
Pd(111), one of the commonly used contact materials recently shown
to exhibit low contact resistance,\cite{Xia2011} as a model to
investigate the influence of thickness and orientation on the work
functions of graphene layers on metals.

Here, we present results of in situ low-energy electron microscopy
(LEEM) and density functional theory (DFT) investigations of the
growth structure and work function of bilayer graphene on Pd(111).
Selected area low-energy electron diffraction (LEED) patterns
indicate that Bernal stacking is typically not observed in the
as-grown graphene layers. From the electron reflectivity data, we
have extracted work functions of graphene domains as a function of
the in-plane orientations of the top and bottom layers. For
monolayer graphene, we have previously shown that the work functions
can vary by up to ~0.15 eV depending on the orientation of the
domains with respect to the substrate.\cite{Murata2010APL} Different
registries between the graphene monolayer and the Pd(111) substrate
lead to strong spatial variations of the charge transfer between the
graphene and substrate. The resulting net surface dipole moment
varies with the in-plane orientation of the monolayer. In case of
bilayer graphene, we find that a smaller net electric dipole
develops between the topmost graphene layer and the rest of the
system ({\em i.e.}, the monolayer-covered substrate). The direction
of this secondary dipole moment is opposite to that of the dipole
developed at the Pd/graphene interface, and leads to an increase in
the work function as compared to monolayer-covered Pd. Furthermore,
the magnitude of the secondary dipole changes with the orientation
of the second layer. This observation suggests that the first
graphene layer does not fully passivate the substrate but allows
charge transfer into the second layer, thus affecting the work
function of the supported bilayer graphene.

\section{Experimental Methods and Results}

\subsection{Methods}
All of our experiments are carried out using a carbon-saturated
Pd(111) single crystal in an ultra-high vacuum (UHV, base pressure
$< 1.0 \times 10^{-10}$ Torr) LEEM system.\cite{Loginova2009,
{McCarty2001}} Sample preparation and other experimental details are
presented in Ref.~\onlinecite{Murata2010APL}. Graphene layers of
desired thickness are obtained by cooling from $\sim$ 900 $^\circ$C
to lower temperatures, during which C segregates from the bulk to
the surface.\cite{Oshima1997,Hamilton1980,{McCarty2009}} For
example, monolayer graphene is obtained when the sample is cooled
from $\sim$ 900 to $\sim$ 790 $^\circ$C at the rate of $\sim$ 1 K/s.
Upon completion of the first layer, further cooling to $\sim$ 700
$^\circ$C or lower yields multilayer graphene. Bright-field LEEM
images are acquired as a function of time at the annealing
temperature. The direct observation of graphene formation helps
identify the layer thickness and is also useful in understanding the
nucleation and growth kinetics. After the growth of a desired number
of layers, the sample is quench-cooled to room temperature. LEEM
images are then acquired as a function of electron energy E at 0.1
eV intervals between $E$ = -5 and 30 eV and 0.01 eV intervals
between $E$ = 0.0 and 2.5 eV. From the LEEM image intensity $I$ vs.
$E$ data, we determine the graphene layer thickness, the electron
injection threshold energy $\phi$, and the work function
$\Phi$.\cite{Babout1980,{babout1977}} Selected area LEED
patterns\cite{Loginova2009} are used to identify the orientation
$\theta$ of graphene layers with respect to the substrate. We define
$\theta$ as the angle between Pd$[1\overline{1}0]$ and graphene$[11
\overline{2}0]$ directions with the positive (negative) sign
denoting in-plane, clockwise (counterclockwise) rotation. The
measurement uncertainties in $\theta$ values are $\pm 1^\circ$.

\subsection{LEEM results}

\begin{figure}[htbp]
\begin{center}
\includegraphics[width=8.0cm]{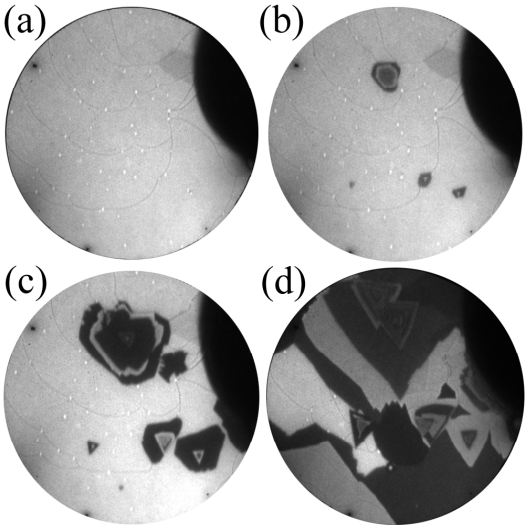}
\caption{Representative LEEM images of Pd(111) acquired after
cooling from 790 $^\circ$C to 660 $^\circ$C during the growth of
multilayer graphene at times $t$ = (a) 17 s, (b) 39 s, (c) 83 s, and
(d) 205 s. ($t$ = 0 corresponds to the time at which the sample
temperature is 660 $^\circ$C.) Field of view = 9.3 $\mu$m and
electron energy $E$ = 4.3 eV. } \label{fig1:LEEM}
\end{center}
\end{figure}

Figure~\ref{fig1:LEEM} shows a series of bright-field LEEM images
acquired from a Pd(111) sample during the growth of multilayer
graphene at 660 $^\circ$C. The sample is cooled from 790 $^\circ$C
when the surface was already covered with 1 ML graphene
(Fig.~\ref{fig1:LEEM}a). The alternating brighter and darker grey
features in Figs.~\ref{fig1:LEEM}b-d, as we show below, are
multilayer stacks of graphene. (The black contrast region observed
along the upper right corner of the LEEM images is due to a
three-dimensional Pd mound on the surface.) We observe three
interesting phenomena: (i) growth occurs at selected locations on
the sample, suggestive of heterogeneous nucleation, presumably at
surface defects; (ii) multiple layers nucleate and grow
simultaneously, {\em i.e.}, third and higher layers appear before
the completion of the second layer. In this measurement sequence,
within 205 s of cooling to 660 $^\circ$C, most of the nucleated
sites are covered with ten or more layers of graphene at the
nucleation sites (see Fig.~\ref{fig1:LEEM}d); and (iii) the
multilayer domains are faceted with regular triangular and/or
truncated hexagonal shapes, indicative of strong anisotropy in step
energies.

\begin{figure}[htbp]
\begin{center}
\includegraphics[width=6.0cm]{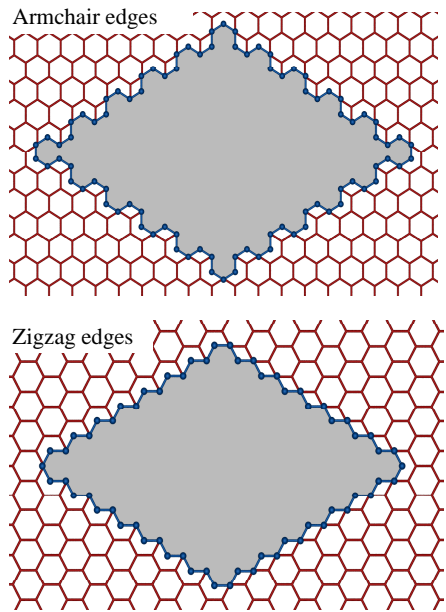}
\caption{Top-view schematic of Bernal-stacked bilayer graphene
lattice. The top and bottom panels show the upper graphene domain
bounded by armchair and zigzag edges, respectively. For clarity, the
lower graphene layer is wider and shows the carbon network, while
the upper layer shows only the edge atoms and their first
neighbors.} \label{fig2:edges}
\end{center}
\end{figure}

We explain this observation as follows. Graphene's honeycomb lattice
gives rise to two types of simple edge structures, armchair and
zigzag. The angle between any two armchair or any two zigzag edges
is 60$^\circ$ or 120$^\circ$, and that between armchair and zigzag
edges is 30$^\circ$ or 90$^\circ$. Therefore, if the edges of the
equilateral triangular and/or truncated hexagonal shapes are simple,
they must all be either zigzag or armchair in structure. By
comparing the directions of domain edges observed in the LEEM images
with the LEED spot orientations, we determined that the domains are
bounded by graphene sheets with zigzag edges. We suggest that
asymmetries in the geometry of zigzag edges relative to an adjacent
graphene layer may account for the anisotropic shapes of multilayer
domains. The asymmetry is most easily illustrated for the case of
two Bernal-stacked layers, as shown in Fig.~\ref{fig2:edges}. All of
the armchair edges are equivalent. However, any two zigzag edges
oriented at 120$^\circ$ to each other are not equivalent, analogous
to the A- and B- type steps on (111)-oriented face-centered cubic
crystals. As we describe in the next section, however, none of the
observed graphene bilayer domains exhibit Bernal stacking (see
Fig.~\ref{fig3:LEEM-multilayerGr}). Yet, it is possible that the
energies of both armchair and zigzag edges can vary with their
orientation relative to the adjacent layer. This may explain the
observation of multiple types of anisotropic shapes.

In the following sections, we report the thickness, crystallographic
orientation, and work functions of the graphene layers.
Fig.~\ref{fig3:LEEM-multilayerGr}a is a representative LEEM image of
multilayer graphene grown by cooling the sample from 900 $^\circ$C
to 709 $^\circ$C and held for 1200 s.
Fig.~\ref{fig3:LEEM-multilayerGr}b is a typical plot of $I$ vs. $E$
data collected from the regions (color coded for clarity)
highlighted in Fig. \ref{fig3:LEEM-multilayerGr}a. Note the
oscillations in $I$ at $E$ values between $\sim$ 1 and $\sim$ 8 eV.
These oscillations are a direct consequence of interference between
electrons reflecting from the surface and those from interlayer
and/or layer-substrate interfaces.\cite{Hibino2008} The number of
oscillations increases with increasing number of layers. For
graphene on Pd(111), we find that $n$-layer graphene, irrespective
of the interlayer stacking, exhibits ($n$-1) oscillations, {\em
i.e.}, 1, 2, 3, and 4 ML show 0, 1, 2, and 3 oscillations,
respectively (Fig.~\ref{fig3:LEEM-multilayerGr}b). This result is
similar to the $I$ vs. $E$ data for multilayer graphene on SiC(0001)
after accounting for the so-called "buffer"
layer.\cite{Hibino2008,{Ohta2008}}

\subsection{LEED results}

\begin{figure}[htbp]
\begin{center}
\includegraphics[width=8.0cm]{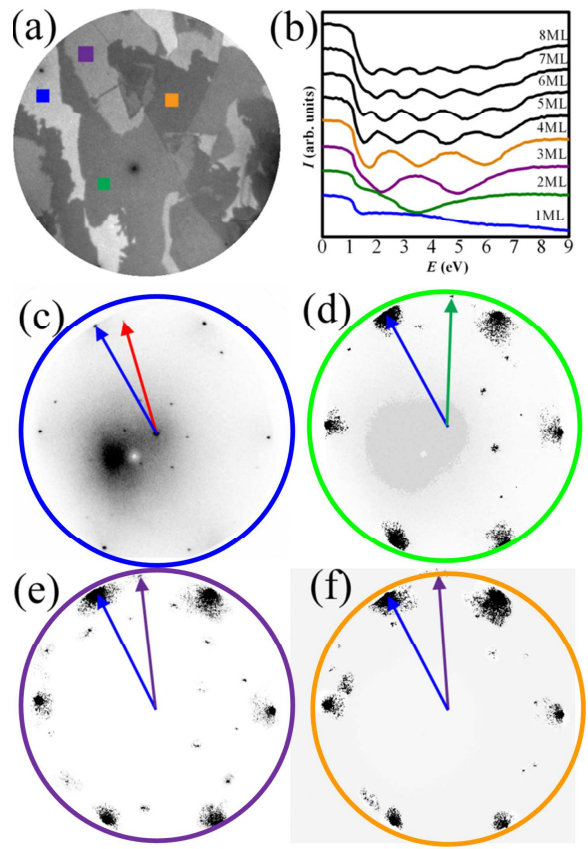}
\caption{(a) Typical LEEM image of multilayer graphene on Pd(111).
Field of view = 9.3 $\mu$m, $E$ = 3.25 eV. (b) Image intensity $I$
vs. $E$ for multilayer graphene. (c-f) Selected area LEED patterns
of 1, 2, 3, and 4 ML graphene-covered regions highlighted by blue,
green, purple, and orange squares in (a), respectively. Red arrow in
(c) indicates the Pd$[1\overline{1}0]$. The blue arrow indicates the
graphene$[11\overline{2}0]$ for the topmost-graphene sheet. The
green and purple arrows indicate graphene$[11\overline{2}0]$ for the
buried graphene sheets that give the second strongest diffraction
from 2 and 3 ML, respectively. } \label{fig3:LEEM-multilayerGr}
\end{center}
\end{figure}

Next, we have determined the crystallographic orientations of the
individual graphene layers using selected area LEED in combination
with $I$ vs. $E$ data. Figs.~\ref{fig3:LEEM-multilayerGr}c-f are
representative LEED patterns acquired from the regions, highlighted
in Fig.~\ref{fig3:LEEM-multilayerGr}a, of 1, 2, 3, and 4 ML graphene
analyzed above. We first identify the diffraction spots
corresponding to Pd(111). Of all the LEED patterns in
Figs.~\ref{fig3:LEEM-multilayerGr}c-f, only
Fig.~\ref{fig3:LEEM-multilayerGr}c shows six-fold symmetric spots
corresponding to the Pd(111)-$1\times1$ lattice. The other set of
spots in Fig.~\ref{fig3:LEEM-multilayerGr}c are due to the monolayer
graphene-$1\times1$ lattice, which is rotated with respect to the
substrate by   $\theta=-14^\circ$. This data serves as the reference
for the identification of LEED spots corresponding to 2, 3, and 4 ML
graphene in Figs.~\ref{fig3:LEEM-multilayerGr}d-f. Since the LEED
spot intensity decreases with increasing distance into the surface,
we can determine the position of the graphene layer with respect to
the surface.

For example, in Fig.~\ref{fig3:LEEM-multilayerGr}d, we observe two
sets of six-fold symmetric spots due to bilayer graphene, with the
stronger and weaker intensity spots oriented along  $ \theta=
-14^\circ$ and $ \theta = +14^\circ$, respectively. That is, the top
graphene layer is oriented along  $\theta_1 = -14^\circ$ while the
layer closer to the substrate is along $\theta_2 = +14^\circ$. The
LEED patterns from 3 and 4 ML graphene, in
Figs.~\ref{fig3:LEEM-multilayerGr}e and
\ref{fig3:LEEM-multilayerGr}f, respectively also show two sets of
six-fold symmetric spots with the strongest spots along $\theta =
-14^\circ$ and the weaker spots along $\theta = +8.9^\circ$. We
identify $\theta= -14^\circ$ as the topmost layer, in both Figs. 3e
and 3f. However, we could not determine which of the underlying
graphene layers gives rise to the $\theta = +8.9^\circ$ spots. In
all of the LEED patterns, we find that one of the layers in
multilayer graphene is oriented at the same angle ( $\theta =
-14^\circ$ in the example of Fig. 3) as in the 1 ML graphene that
surrounds the multilayer stack. This observation suggests that
orientation of the top layer graphene is unaffected during the
growth of subsequent layers, consistent with the expected growth of
multiple layers below the layer that grows first.\cite{Nie2011} From
LEED patterns acquired from over 10 different regions of the sample,
we identified at least 12 different in-plane orientations in bilayer
graphene. In all cases the two layers are not rotationally aligned,
as required for Bernal stacking, similar to previous
reports.\cite{Murata2010APL} Based upon our data, we suggest that
Bernal stacking will be rare in graphene layers grown on Pd(111).

\subsection{Work function of graphene bilayers on Pd}
We  now focus on the relationship between work function $\Phi$ and
orientation of bilayer graphene. Following the procedure outlined in
Ref.~\onlinecite{Murata2010APL}, we first measure the electron
injection threshold energy $\phi$, at which the image intensity $I$
drops to 90\% of the intensity value at $E$ = 0 eV. Then, $\Phi =
\Phi_{fil} + \phi$, where $\Phi_{fil}$ is the work function of the
electron gun filament. From the $I$ vs. $E$ data obtained from clean
Pd(111) along with known values of $\Phi$ for Pd (5.3$-$5.6
eV),\cite{Murata2010APL} we estimate $\Phi_{fil}$ to be between 3.1
and 3.4 eV. Given the large uncertainties in the knowledge of
$\Phi_{fil}$ (up to 0.3 eV), we do not emphasize determining
absolute values of $\Phi$. However, $\phi$ values are precise to
within 0.02 eV and, therefore, can be used to compare how $\Phi$
changes with orientation and layer thickness. Recently, we reported
that $\Phi$ of monolayer graphene on Pd(111) varies with in-plane
rotation due to spatial variations in charge transfer at graphene -
Pd interface.\cite{Murata2010APL} Using the same approach and
assuming that $\Phi_{fil}$ = 3.4 eV, we have determined $Phi$ for
bilayer graphene domains as a function of $\theta_1$ and $\theta_2$.

\begin{figure}
\begin{center}
\includegraphics[width=6.5cm]{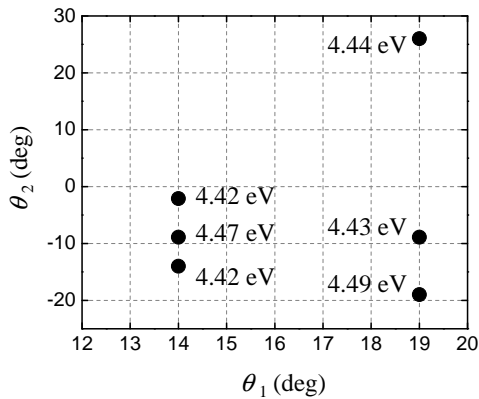}
\caption{Experimental work function values for bilayer graphene on
Pd(111) for different orientations $\theta_1$ and  $\theta_2$ of the
two graphene layers. } \label{fig4:theta-theta-WF}
\end{center}
\end{figure}

Fig. 4 shows the work function values $\Phi$ for bilayer graphene
plotted at different values of $\theta_1$ and $\theta_2$. For all
domain orientations, the work function of bilayer graphene is higher
than that for monolayer graphene. Our results for $\Phi$ of graphene
bilayers on Pd(111) are similar to recent reports on epitaxial
graphene grown on SiC (where the $\Phi$ of bilayer graphene is also
higher than that of a monolayer),\cite{Filleter2008} but opposite to
those for free-standing graphene transferred onto
SiO$_2$,\cite{Lee2009} where $\Phi$ is found to decrease with
increasing layer thickness. We note, however, that the effect of
domain orientation on the $\Phi$ values of bilayer graphene on
Pd(111) is smaller (up to 0.07 eV) compared to that observed (up to
0.15 eV) in $\Phi$ for monolayer graphene domains (see Fig. 4). In
order to confirm these results, we determined $\Phi$ values using a
different procedure,\cite{babout1977,{Babout1980}} and obtained
consistent results. An important factor that affects the work
function of bilayer graphene is the in-plane orientation of the two
layers, which we next discuss from the perspective of electronic
structure calculations.

\section{DFT results and discussion}

In order to gain insights into how the second graphene layer and the
orientation of both layers affect the work function, we have
performed DFT calculations on systems with one and two graphene
layers placed on one side of a Pd(111) substrate. We have used the
Siesta code\cite{Soler2002} with double-zeta (DZ) basis set within
the local density approximation (LDA) with the Ceperley-Alder
exchange-correlation functional. A uniform grid in real space with a
mesh cutoff of 150 Ry was used. The Brillouin zone has been sampled
only at the $\Gamma$ point since the unit cells were large, with
periodicity vectors of 19.72\AA. The residual forces on any atom are
smaller than 0.04 eV/\AA\  at the end of structural relaxations, and
the total energy has been converged to within $10^{-5}$ eV for all
electronic property calculations. We have used $8\times8$ surface
supercells for the construction of 19.9$^\circ$ and
30$^\circ$-rotated graphene domains (126 C atoms and 128 C atoms,
respectively), and 46 Pd atoms per layer with four Pd layers in the
substrate. For the 30$^\circ$ orientation, we have adopted the
epitaxial structure proposed by Giovanetti {\em et
al.}\cite{Giovannetti2008} The choice of the two particular
orientations, 19.9$^\circ$ and 30$^\circ$, is determined both by the
necessity to keep the computations tractable and by the fact that
these two orientations can be placed simultaneously on the Pd(111)
substrate with virtually no strain in either of them. Thus, any
computed variations in $\Phi$ are due only to their orientations
and/or stacking order, but not to the (insignificant) strain in the
graphene layers.

\begin{table*}
\caption{\label{tab:1}Work function results from DFT calculations
for single- and bi-layer graphene, along with the net dipole $P_a$
and the secondary dipole $P_b$ defined in text. The work function of
pure Pd substrate is also given as a reference. The electronic
transfer to the graphene layer(s) is given in the last column, where
1$^{\rm st}$ layer denotes the one closest to the Pd substrate.}
\begin{ruledtabular}
\begin{tabular}{lcccc}
System  & Work      & Net dipole        & Secondary     & Electrons \\
        & Function  & $P_a$         & dipole, $P_b$ & transferred (e/\AA$^2$) \\
        & (eV)      & (D/\AA$^2$)    & (D/\AA$^2$)    & 1$^{\rm st}$ layer [2$^{\rm nd}$ layer] \\
\hline
Pd                              & 5.165 & ---       & ---       & --- \\
30.0$^\circ$/Pd                 & 4.131 & +0.027    & ---       & -0.0027 \\
19.9$^\circ$/Pd                 & 4.082 & +0.027    & ---       & -0.0023 \\
30.0$^\circ$/19.9$^\circ$/Pd    & 4.307 & +0.019    & -0.001    & -0.0029 [+0.0003] \\
19.9$^\circ$/30.0$^\circ$/Pd    & 4.738 & +0.011    & -0.014    & -0.0056 [+0.0027] \\
30.0$^\circ$/30.0$^\circ$/Pd    & 4.337 & +0.019    & -0.003    & -0.0041 [+0.0009] \\
19.9$^\circ$/19.9$^\circ$/Pd    & 4.723 & +0.008    & -0.012    & -0.0048 [+0.0023] \\
\end{tabular}
\end{ruledtabular}
\end{table*}

\begin{figure}[htbp]
\begin{center}
\includegraphics[width=9.0cm]{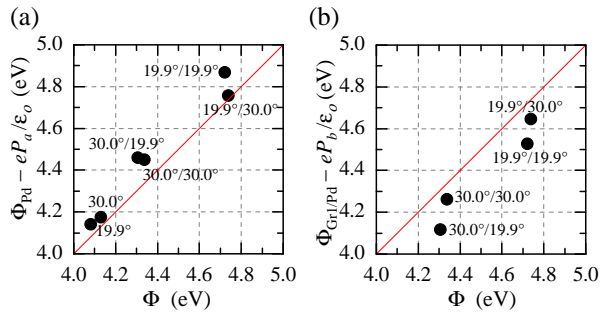}
\caption{Work function values from dipole moment calculations,
$\Phi_{ref} - eP/\epsilon_0$, vs. values determined directly from
DFT calculations, $\Phi$. (a) Bare Pd surface is taken as the
reference for single- and bi-layer epitaxial systems. (b) Pd
substrate covered with monolayer graphene is used as reference for
the bi-layer calculations.} \label{fig5:phi-phi-plots}
\end{center}
\end{figure}

\begin{figure*}[htbp]
\begin{center}
\includegraphics[width=12.0cm]{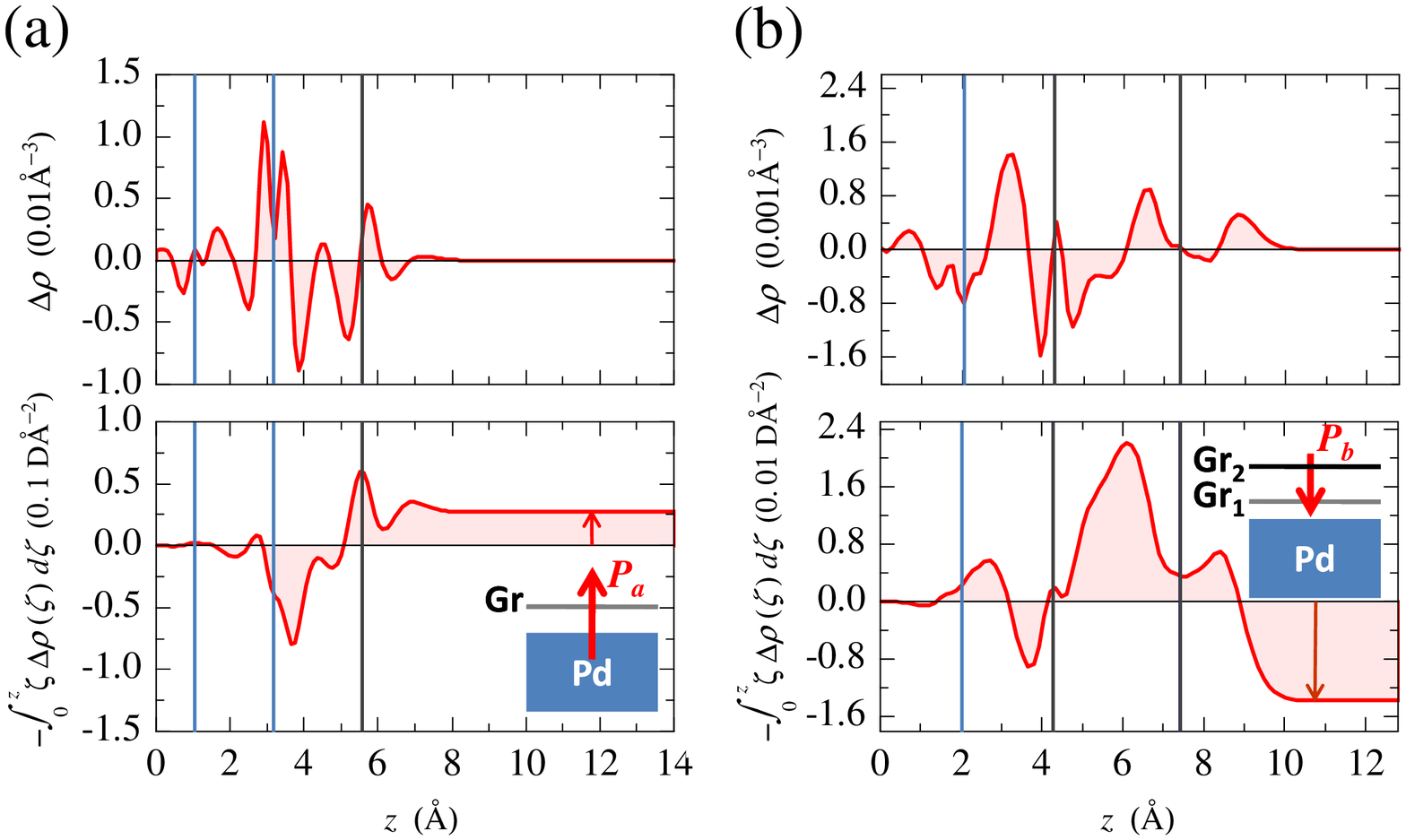}
\caption{Plane-averaged electron density transferred    (number of
electrons per unit volume) and surface dipole densities for (a)
30.0$^\circ$/Pd and (b) 19.9$^\circ$/30.0$^\circ$/Pd. The vertical
gray (blue) lines indicate the locations of the graphene (substrate)
layers. The origin of the $z$ coordinate is taken in the middle of
the corresponding reference substrates, {\em i.e.}, (a) the midpoint
of the bare Pd slab, and (b) the midpoint of the segment between
inner (first) graphene layer and bare Pd layer on the opposite side
of the slab. The arrows in the insets show schematically (a) the
surface dipole along the surface normal for monolayer graphene, and
(b) the secondary dipole pointing opposite to the surface normal. }
\label{fig6:chargeANDdipole}
\end{center}
\end{figure*}

For monolayer graphene, we compute the charge transfer and the
associated dipole moment in a manner similar to previous
reports,\cite{Wang2010} but strictly adopt the standard direction
convention for electrostatic dipole ({\em i.e.}, dipole vector
pointing from the negative charge to the positive charge). For
single- and bi-layer graphene, the planar averaged electron density
transferred with respect to all the individual components
(substrate, first layer, and second layer --when present), is
defined by
\begin{equation}
\Delta \rho_a (z) = \rho_{{\rm Gr}_2/{\rm Gr}_1/{\rm Pd}} (z) -
\rho_{\rm Pd} (z) -\rho_{{\rm Gr}_1}(z)-\rho_{{\rm Gr}_2} (z).
\end{equation}
This electron transfer creates a net dipole moment when integrated
from the middle of the substrate to the middle of the vacuum in the
supercell, $P_a = \int z \Delta \rho_a (z) ${\rm d}$z$, dipole which
should be responsible for the change in $\Phi$ with respect to the
bare Pd surface
\begin{equation}
 \Phi_{{\rm Gr}_2/{\rm Gr}_1/{\rm Pd}} = \Phi_{\rm Pd} - eP_a/\epsilon_0 .
\end{equation}
Indeed, computing the work function directly from the electrostatic
potential output of the DFT simulation (= $\Phi_{\rm {vacuum}} -
E_{\rm {Fermi}}$) and from the dipole moment $P_a$ [via Eq.~(2)]
shows that the electrostatic dipole  largely accounts for the work
function of single- and bi-layer graphene on Pd (Fig. 5a). For both
single- and bi-layer graphene, the values of the dipole moment $P_a$
are positive ({\em i.e.}, pointing away from the substrate, see
Table 1) and of same order of magnitude as that reported for other
graphene-metal systems.\cite{Wang2010}

To account specifically for the influence of the second graphene
layer on $\Phi$, we resort to another way of computing the dipole
moment, one in which the reference consists of the topmost (second)
graphene layer and the palladium substrate covered with monolayer
graphene,
\begin{equation}
\Delta \rho_b (z) = \rho_{{\rm Gr}_2/{\rm Gr}_1/{\rm Pd}} (z) -
\rho_{{\rm Gr}_1/{\rm Pd}} (z)-\rho_{{\rm Gr}_2} (z).
\end{equation}
The charge transfer $\Delta \rho_b (z)$ and the dipole moment $P_b =
\int z \Delta \rho_b (z) ${\rm d}$z$ developed with respect to the
new reference are, therefore, much smaller when compared to the
transfer to the first layer, as shown in Fig. 6 and listed in Table
1. Although smaller, the secondary dipole moment $P_b$ does account
for the work function changes between the single- and bi- layer
graphene on Pd,
\begin{equation}
\Phi_{{\rm Gr}_2/{\rm Gr}_1/{\rm Pd}}=\Phi_{{\rm Gr}_1/{\rm
Pd}}-eP_b/\epsilon_0,
\end{equation}
as seen in Fig. 4b. The secondary dipole moment $P_b$ (Table 1) is
always negative ({\em i.e.}, it points into the substrate), which
explains our observation that the work functions of bilayer systems
are greater than those of monolayer systems. We note that although
the variations in experimentally measured values are in qualitative
agreement with the DFT variations, they are smaller than those
obtained using DFT. Possible reasons for the smaller variations in
experimental values are (a) different orientations used in
calculations compared to experiments (our initial attempts to use
the experimental orientations in DFT simulations resulted in
supercells that were either too large or too strained), and (b) the
presence of elastic strain, carbon adatoms at the graphene-Pd
interface, or other impurities in graphene layers.

Finally, we have also estimated the electron transfer to each
graphene layer by integrating the density using as integration
limits the midpoints between atomic layers. For computing the
electron transfer to the second graphene layer (topmost), we
integrated from the midpoint between the first and second graphene
layers to the middle of the vacuum. The results, listed in the last
column of Table 1, show that the first graphene layer is $p$-doped
(loss of electrons), while the second layer is $n$-doped. As seen in
Table 1, the magnitude of the electron transfer to the second layer
is relatively small and can be attributed to charge screening by the
first layer.\cite{Sun2010} The signs of the estimated electronic
transfer to each graphene layer are certainly consistent with
directions of the calculated dipoles $P_a$ and $P_b$, {\em i.e.}, a
loss of electrons on the first graphene layer leads to a dipole
$P_a$ parallel to the surface normal, while a gain of electrons on
the second graphene layers leads to a secondary dipole $P_b$
antiparallel to the surface normal.

\section{Conclusions}

In conclusion, we used in {\em situ} LEEM and DFT to investigate the
growth structure and work function of multilayer graphene on
Pd(111). None of the bilayer domains we examined had the rotational
alignment required for Bernal stacking. Instead, there are multiple
orientations of the layers relative to the substrate and each other.
The orientation-dependent variations in work functions of bilayer
graphene are relatively smaller than those observed in monolayer
graphene. We explain these results using DFT calculations, which
reveal that the charge transfer from Pd depends sensitively on the
orientations of both graphene layers. Our results indicate that one
cannot infer how multilayers of graphene interact with a metal
substrate (for example, in a contact) by examining only the first
layer. Therefore, for the fabrication of graphene-based transistors
and/or transparent conductors, knowledge of the graphene layer
stacking and its influence on the device characteristics is
desirable.

{\em Acknowledgements.}Sandia work was supported by the Office of
Basic Energy Sciences, Division of Materials Sciences and
Engineering of the US DOE under Contract No. DE-AC04-94AL85000. We
gratefully acknowledge support from UC COR-FRG and from the NSF
through Grants No. OCI-1048586, CMMI-0825592 and CMMI-0846858. We
thank N.C. Bartelt for valuable discussions and comments.


\begin{thebibliography}{99}


\bibitem{Geim2007}A. K. Geim and K. S. Novoselov, Nature Materials 6, 183 (2007).

\bibitem{Kim2009} K. S. Kim, Y. Zhao, H. Jang, S. Y. Lee, J. M. Kim, J. H. Ahn, P.
Kim, J. Y. Choi and B. H. Hong, Nature 457, 706 (2009).

\bibitem{Bae2010} S. Bae, H. Kim, Y. Lee, X. F. Xu, J. S. Park, Y. Zheng, J.
Balakrishnan, T. Lei, H. R. Kim, Y. I. Song, Y. J. Kim, K. S. Kim,
B. Ozyilmaz, J. H. Ahn, B. H. Hong and S. Iijima, Nature
Nanotechnology 5, 574 (2010).

\bibitem{Zhang2009}  Y. B. Zhang, T. T. Tang, C. Girit, Z. Hao, M. C. Martin, A.
Zettl, M. F. Crommie, Y. R. Shen and F. Wang, Nature 459, 820
(2009).

\bibitem{Xia2011} F. N. Xia, V. Perebeinos, Y. M. Lin, Y. Q. Wu and P. Avouris,
Nature Nanotechnology 6, 179 (2011).

\bibitem{Kwon2009}  S. Y. Kwon, C. V. Ciobanu, V. Petrova, V. B. Shenoy, J. Bareno,
V. Gambin, I. Petrov and S. Kodambaka, Nano Letters 9, 3985 (2009).

\bibitem{Murata2010APL}  Y. Murata, E. Starodub, B. B. Kappes, C. V. Ciobanu, N. C.
Bartelt, K. F. McCarty and S. Kodambaka, Applied Physics Letters 97,
143114 (2010).

\bibitem{Murata2010ACSNano} Y. Murata, V. Petrova, B.B. Kappes, A. Ebnonnasir, I. Petrov,
Y.-H. Xie, C.V. Ciobanu, and S. Kodambaka, ACS Nano 4, 6509 (2010)

\bibitem{Loginova2009} E. Loginova, N. C. Bartelt, P. J. Feibelman and K. F. McCarty,
New Journal of Physics 11, 063046 (2009).

\bibitem{McCarty2001} K. F. McCarty, Surface Science 474, L165 (2001).

\bibitem{Oshima1997} C. Oshima and A. Nagashima, Journal of Physics-Condensed Matter
9, 1 (1997), and references therein.

\bibitem{Hamilton1980} J. C. Hamilton and J. M. Blakely, Surface Science 91, 199 (1980).

\bibitem{McCarty2009} K. F. McCarty, P. J. Feibelman, E. Loginova and N. C. Bartelt, Carbon 47, 1806 (2009).

\bibitem{Babout1980} M. Babout, M. Guivarch, R. Pantel, M. Bujor and C. Guittard,
Journal of Physics D-Applied Physics 13, 1161 (1980).

\bibitem{babout1977} M. Babout, J. C. Lebosse, J. Lopez, R. Gauthier and C. Guittard,
Journal of Physics D-Applied Physics 10, 2331 (1977).

\bibitem{Hibino2008} H. Hibino, H. Kageshima, F. Maeda, M. Nagase, Y. Kobayashi and
H. Yamaguchi, Physical Review B 77, 075413 (2008).

\bibitem{Ohta2008} T. Ohta, F. El Gabaly, A. Bostwick, J. L. McChesney, K. V.
Emtsev, A. K. Schmid, T. Seyller, K. Horn and E. Rotenberg, New
Journal of Physics 10, 023034 (2008).

\bibitem{Nie2011} S. Nie, A. L. Walter, N. C. Bartelt, E. Starodub, A. Bostwick,
E. Rotenberg and K. F. McCarty, ACS Nano 5, 2298 (2011).

\bibitem{Filleter2008} T. Filleter, K. V. Emtsev, T. Seyller and R. Bennewitz, Applied
Physics Letters 93, 133117 (2008).

\bibitem{Lee2009} N. J. Lee, J. W. Yoo, Y. J. Choi, C. J. Kang, D. Y. Jeon, D. C.
Kim, S. Seo and H. J. Chung, Applied Physics Letters 95, 222107
(2009).

\bibitem{Soler2002} J. M. Soler, E. Artacho, J. D. Gale, A. Garcia, J. Junquera, P.
Ordejon and D. Sanchez-Portal, Journal of Physics-Condensed Matter
14, 2745 (2002).

\bibitem{Giovannetti2008} G. Giovannetti, P. A. Khomyakov, G. Brocks, V. M. Karpan, J. van
den Brink and P. J. Kelly, Physical Review Letters 101, 026803
(2008).

\bibitem{Wang2010} B. Wang, S. G$\ddot{u}$nther, J. Wintterlin and M. L. Bocquet, New
Journal of Physics 12, 043041 (2010).

\bibitem{Sun2010} D. Sun, C. Divin, C. Berger, W. A. de Heer, P. N. First and T.
B. Norris, Physical Review Letters 104, 136802 (2010).



\end{thebibliography}
\end{document}